\documentstyle[prd,aps,floats,epsfig,times]{revtex}
\draft
\begin{document}
\title{On the absence of scalar hair for charged rotating blackholes in non minimally coupled theories}
\author{S. Sen$^{\star}$ and N. Banerjee$^{\dagger}$}
\address{$^{\star}$ Harish Chandra Research Institute, Chhatnag Road, Jhusi, Allahabad 211 019, India. E-mail: somasri@mri.ernet.in\\
$^{\dagger}$ Department of Physics, Jadavpur University, Calcutta- 700 032, India.E-mail: narayan@juphys.ernet.in}
\maketitle
\begin{abstract}
In this work we check the validity of the no scalar hair 
theorem in charged axisymmetric stationary black holes for 
a wide class of scalar tensor theories. \\

{\bf keywords:} Blackhole; no scalar hair; axisymmetric spacetime. 
\end{abstract}



\section{Introduction}
\vspace*{-0.5pt}
\noindent
Recently there has been a considerable resurgence in the no scalar hair 
theorem for black holes. Investigations regarding no hair theorem, however, 
had started about thirty years ago\cite{R1}.
Inspired by Israel's uniqueness theorem for Schwarzschild and
Reissner-Nordstrom black holes\cite{R2} and Carter\cite{R3} and Wald's\cite{R4} uniqueness theorem
for Kerr black holes, Wheeler anticipated that gravitational collapse leads to black
holes endowed with mass, charge and angular momentum and no other free
parameters, which he summerised as `black holes have no hair'. 
The `no scalar hair theorem' excludes 
the availability of any knowledge of a scalar field from the exterior geometry
of a black hole even when a scalar field is present in the 
spacetime along with gravity.
\par
The search for such scalar hair were initiated long back.
 Investigations, involving
physical fields like massless scalar\cite{R5},
massive vector\cite{R6}, spinor\cite{R7} fields go in favour
 of Wheeler's dictum as any
information about these fields from a stationary\cite {R6} black hole
exterior is excluded. 
These investigations were mainly limited to the cases where the scalar
field is only minimally coupled to gravity.
But in the early 90's, solutions for
stationary black holes with exterior non-abelian gauge field or skyrmion
field\cite{R8,R9,R10} have put strong challenge in front of the conjecture. Black hole
solutions with new hair like Yang Mills hair\cite{R8}, Skyrme hair\cite{R9}, dilaton hair\cite{R11} or
others\cite{R12} act as counter examples to the conjecture. With a few exceptions\cite{R9} many
of these black holes are unstable\cite{R13}. It is interesting to note that all the hair are not of similar stature\cite{R14}. The
hair which act as new quantum numbers, i.e, independent of other quantum
numbers are primary hair. Skyrme hair\cite{R9,R14,R15} in nonlinear sigma models 
coupled to 
gravity are examples of such hair. The hair which grow on other hair,
i.e, the new quantum numbers determined by other quantum numbers are
examples of secondary hair. Dilaton hair on electrically charged black
holes\cite{R11},Kaluza Klein black holes\cite{R16}  fall in this second category.
\par 
In spite of the popular name, there is no proof of the no hair
theorem, and its status is in fact that of a conjecture. In the
absence of a true theorem, one has to consider explicitly various sources of gravity and try
to examine the nature of admissible black hole solutions. In this work the
validity of the no scalar hair theorem is studied for a class of
stationary axisymmetric charged black hole solutions in the context of a
wide class of scalar tensor theories. In the rotating spacetime there were
some investigations\cite{R6},\cite{R17} with minimally coupled scalar fields. 
It was
showed that black hole in its final state cannot be endowed 
with an exterior scalar field.
So the interest in the present work
primarily involves the inclusion of a wide class of scalar
tensor theories in axially symmetric spacetime, 
where the scalar field is nonminimally coupled to gravity. 
\par
In order to check the validity of the no scalar hair theorem we have
explicitly studied the spacetime metric and scalars both in the cases of
minimally and nonminimally coupled scalar fields. But 
the exact solutions for such fields in various scalar tensor theories are
not available in the literature in most cases. So we use an algorithm 
to generate 
the exact solutions for charged rotating spacetime with a minimally 
coupled scalar field from the known general relativistic electrovac solution. 
This solution is
then analysed to find the compatibility of a scalar field with a black
hole. Then we use a conformal transformation to
generate the solutions for a large number of non minimally coupled scalar
tensor theories  from  Einstein
Maxwell minimally coupled scalar field (EMS) solution and test 
the no scalar hair theorem against these solutions.

\section{A minimally coupled scalar field}
\subsection{A technique to generate solutions for minimally coupled scalar
field}
We start with a general form of stationary axially symmetric line element
$$
ds^{2}=e^{2\psi}(dt+\omega d\phi)^{2}-e^{-2\psi}[e^{2\gamma}(dx_{1}^{2}+dx_{2}^{2})+h^{2}d\phi^{2}],
\eqno{(2.1)}$$
where $\psi, \omega, \gamma, h$ are all functions of $x_{1}$ and $x_{2}$.\\
The energy momentum tensor for the electromagnetic field is
$$
T_{\mu\nu}=g^{\alpha\beta}F_{\mu\alpha}F_{\nu\beta}-{1\over{4}}g_{\mu\nu}F_{\alpha\beta}F^{\alpha\beta},
\eqno{(2.2)}$$
where the Maxwell tensor $F_{\mu\nu}$ is given by 
$$
F_{\mu\nu}=A_{\nu,\mu}-A_{\mu,\nu},
\eqno{(2.3)}$$
$A_{\mu}$ being the vector potential component. $A_{0}$ and $A_{3}$, (i.e $A_{t}$ and 
 $A_{\phi}$) are 
the only existing components of $A_{\mu}$. They are also functions of $x_{1}$ and $x_{2}$.
\par
Now if a massless scalar field $\phi$ is also included, the total energy momentum 
tensor becomes
$$
E_{\mu\nu}=T_{\mu\nu}+S_{\mu\nu},
\eqno{(2.4)}$$
where $T_{\mu\nu}=$ energy momentum tensor for electromagnetic field\\
and $S_{\mu\nu}=$ energy momentum tensor due to massless scalar field \\
$\hspace{15mm}=\phi_{,\mu}\phi_{,\nu}-{1\over{2}}g_{\mu\nu}\phi_{,\alpha}\phi^{,\alpha}$, where $\phi$ is also function of $x_{1}$ and $x_{2}$.
\par
The set of equations to be solved are
$$
R_{\mu\nu}=-\phi_{,\mu}\phi_{,\nu}-g^{\alpha\beta}F_{\mu\alpha}F_{\nu\beta}+{1\over{4}}
g_{\mu\nu}F_{\alpha\beta}F^{\alpha\beta},
\eqno{(2.5)}$$
$$\Box\phi=0,\eqno{(2.6)}$$
and $$F^{\mu\nu}_{;\nu}=0.\eqno{(2.7)}$$
 For the line element (2.1), the wave
equation (2.6) becomes
$$\phi_{11}+\phi_{22}+{h_{1}\over{h}}\phi_{1}+{h_{2}\over{h}}\phi_{2}=0.
\eqno{(2.8)}$$
Now according to the generation technique, if $\gamma$ can be written as 
$$\gamma=\gamma^{v}+\gamma^{\phi},\eqno{(2.9)}$$
where $\gamma^{v}$ is the solution for $\gamma$ in the electrovac field 
for metric (2.1) and $\gamma^{\phi}$ satisfies the equations 
$$
h_{1}\gamma_{2}^{\phi}+h_{2}\gamma^{\phi}_{1}=h\phi_{1}\phi_{2},
\eqno{(2.10)}$$
$$
h_{1}\gamma_{1}^{\phi}-h_{2}\gamma^{\phi}_{2}={h\over{2}}(\phi_{1}^{2}-\phi_{2}^{2}),
\eqno{(2.11)}$$
then, the metric coefficients $\psi, \omega$ and $h,$ vector potentials $A_{0}$ and
$A_{3}$ of the general relativistic electrovac solutions along with $\gamma$ as given by (2.9) and $\phi$
 determined by (2.8) form the complete set of solutions for Einstein-Maxwell field 
 minimally coupled with massless scalar field (EMS).
\par
This algorithm is similar to that given by Eris and Gurses\cite{R18}. The difference is that our technique 
holds for a general metric while 
they\cite{R18} have used a 
different coordinate system, where
the metric (2.1) is written in the Weyl Papapetrou canonical form
$$
ds^{2}=e^{2\psi}(dt+\omega d\phi)^{2}-e^{-2\psi}[e^{2\gamma}(d\rho^{2}+dz^{2})+\rho^{2}d\phi^{2}], 
\eqno{(2.12)}$$
$\rho$ and $z$ are harmonic functions of $x_{1}$ and $x_{2}$ and are
called canonical cylindrical coordinates. The
result of reference\cite{R18} can be recovered from our result for
$h=\rho$. 

\subsection{Some axisymmetric solutions with minimally coupled scalar field}
The most widely used 
axially symmetric stationary electrovac solution in general relativity is 
 the Kerr-Newman (KN) metric. We use this solution as a seed for
the algorithm described above.
The KN solution in the well known Boyer Lindquist form is given by
$$
ds^{2}=dt^{2}-{2mr-e^{2}\over{r^{2}+a^{2}\cos^{2}\theta}}(dt+a\sin^{2}\theta d\phi)^{2}
-(r^{2}+a^{2}\cos^2{\theta})\left\{d\theta^{2}+{dr^{2}\over{r^{2}-2mr+a^{2}+e^{2}}}\right\}$$
$$\hspace{33mm}
-(r^{2}+a^{2})\sin^{2}\theta d\phi^{2},\eqno{(2.13)}$$
and the solutions for the vector potentials are
$$A_{3}=-{ear\sin^{2}\theta\over{r^{2}+a^{2}\cos^{2}\theta}}\hspace{10mm}and\hspace{10mm}
A_{0}=-{er\over{r^{2}+a^{2}\cos^{2}\theta}}.\eqno{(2.14)}$$
The constants $m,a$ and $e$ are the mass, angular momentum per unit mass
and the electric charge respectively of the axisymmetric distribution.
\par
By a coordinate transformation of the form
$$r=e^{R}+m+{m^{2}-a^{2}-e^{2}\over{4}}e^{-R},\eqno{(2.15)}$$
Misra et al\cite{R20} and later Singh et al\cite{R21} had rewritten the KN metric 
in the following canonical form
$$
ds^{2}=\left({L^{2}-2mL+a^{2}\cos^{2}\theta+e^{2}\over{L^{2}+a^{2}\cos^{2}\theta}}\right)
\left[dt-{(2mL-e^{2})a\sin^{2}\theta\over{L^{2}-2mL+a^{2}\cos^{2}\theta+e^{2}}}d\phi\right]^{2}$$
$$
-\left({L^{2}+a^{2}\cos^{2}\theta\over{L^{2}-2mL+a^{2}\cos^{2}\theta+e^{2}}}\right)
\{(L^{2}-2mL+a^{2}\cos^{2}\theta+e^{2})(dR^{2}+d\theta^{2})$$
$$\hspace{15mm}
+(L^{2}-2mL+a^{2}+e^{2})\sin^{2}\theta d\phi^{2}\}
\eqno{(2.16)}$$
where
$$ L=e^{R}+m+{m^{2}-a^{2}-e^{2}\over{4}}e^{-R}.\eqno{(2.17)}$$ 
The vector potentials in the transformed coordinates are
$$A_{3}=-{eaL\sin^{2}\theta\over{L^{2}+a^{2}\cos^{2}\theta}}\hspace{10mm}and\hspace{10mm}
A_{0}=-{eL\over{L^{2}+a^{2}\cos^{2}\theta}}.\eqno{(2.18)}$$
We shall now use this solution to generate the corresponding EMS solution.
 In terms of the metric
(2.16), equation (2.8) can be written as
$$\phi_{RR}+\phi_{\theta\theta}+{e^{R}+{M^{2}\over{4}}e^{-R}\over{e^{R}-{M^{2}\over{4}}e^{-R}}}
\phi_{R}+{\cos\theta\over{\sin\theta}}\phi_{\theta}=0,
\eqno{(2.19)}$$
where $$
M^{2}=m^{2}-a^{2}-e^{2}.\eqno{(2.20)}$$
Equation (2.19) can be solved in a general way by simply imposing the separability condition. The scalar field $\phi$ is considered to be separable in
functions of $R$ and $\theta$ both in product and summed
form. The solutions for $\phi$ in both the ways have been exhibited
explicitly in the Appendix. Here we would consider some special cases of
the general solutions.

\subsubsection{First set}
We first consider the simplest case of $\phi[\phi=\phi(R)]$, i.e, 
$\phi$ is isotropic. The solution for $\phi$ in such
case, 
$$
\phi=\phi_{0}+{\sigma\over{M}}\ln\left({e^{R}-{M\over{2}}\over{e^{R}+{M\over{2}}}}\right)=\phi_{0}+{\sigma\over{2M}}\ln{L-(m+M)\over{L-(m-M)}},
\eqno{(2.21)}$$
where $\sigma$ and $\phi_{0}$ are two constants. We take $\phi_{0}=0$
without any loss of generality.
\par
Once $\phi$ is specified, we can find $\gamma^{\phi}$ from equations (2.10) and
(2.11) as
$$\gamma^{\phi}={\sigma^{2}\over{4M^{2}}}\ln {(e^{R}-{M^{2}\over{4}}e^{-R})^{2}\over
{(e^{R}-{M^{2}\over{4}}e^{-R})^{2}+M^{2}\sin^{2}\theta}}$$
$$\hspace{20mm}={\sigma^{2}\over{4M^{2}}}\ln{L^{2}-2mL+a^{2}+e^{2}\over{(L-m)^{2}
-M^{2}\cos^{2}\theta}}.\eqno{(2.22)}$$
Hence
$$
e^{2\gamma}=e^{2\gamma^{v}+2\gamma^{\phi}}=(L^{2}-2mL+a^{2}\cos^{2}\theta+e^{2})
\left\{{L^{2}-2mL+a^{2}+e^{2}\over{(L-m)^{2}-M^{2}\cos^{2}\theta}}\right\}^{\sigma^{2}\over{2M^{2}}}.
\eqno{(2.23)}$$
By using the inverse transformation given by (2.15), the line element in EMS field can be written in the well known Boyer Lindquist form, as
$$
ds^{2}=dt^{2}-{2mr-e^{2}\over{r^{2}+a^{2}\cos^{2}\theta}}(dt+a\sin^{2}\theta d\phi)^{2}
-(r^{2}+a^{2}\cos^{2}\theta)\left\{{r^{2}-2mr+a^{2}+e^{2}\over{(r-m)^{2}-M^{2}\cos^{2}\theta}}\right\}^{\sigma^{2}\over{2M^{2}}}
$$
$$
\left\{d\theta^{2}+{dr^{2}\over{r^{2}-2mr+a^{2}+e^{2}}}\right\}
-(r^{2}+a^{2})\sin^{2}\theta d\phi^{2},\eqno{(2.24)}$$
with the vector potentials 
$$A_{3}=-{ear\sin^{2}\theta\over{r^{2}+a^{2}\cos^{2}\theta}}\hspace{10mm}and\hspace{10mm}
A_{0}=-{er\over{r^{2}+a^{2}\cos^{2}\theta}},\eqno{(2.25)}$$
and the scalar field 
$$
\phi={\sigma\over{2M}}\ln{r-(m+M)\over{r-(m-M)}}.\eqno{(2.26)}$$ 
 The scalar field vanishes 
for the limit $r\rightarrow\infty$ and the metric is asymptotically flat.
 In this solution, for $\sigma=0$, the scalar field 
becomes trivial and the metric goes over to KN solution. If we put off the
electric charge,  
i.e $e=0$, and set the angular momentum also to zero, i.e $a=0$, the metric
 reduces to one of the solutions given by Penny\cite{R22}. 
Without the electric charge, the solution (2.24) reduces to the
Brans-Dicke-Kerr solution given by McIntosh\cite{R23} in Dicke's
revised units\cite{R24}.
\par
From the metric (2.24) we see that $g_{11}$ is singular at $r=m\pm M$ surfaces.
Simultaneously the scalar field $\phi$ in (2.26) diverges at these two surfaces
if $\sigma\neq{0}$. The  Ricci scalar is given by
$$
{\cal{R}}=-\phi_{\alpha}\phi^{\alpha}={\sigma^{2}\over{[r^{2}-2mr+a^{2}+e^{2}]^{1+{\sigma^{2}\over{2M^{2}}}}}}\times
{((r-m)^{2}-M^{2}\cos^{2}\theta)^{\sigma^{2}\over{2M^{2}}}\over{r^{2}+a^{2}\cos^{2}\theta}}.
\eqno{(2.27)}$$
It is evident from this expression that Ricci scalar diverges at the surfaces
$r=m\pm M$ for $\sigma\neq{0}$ and thus these surfaces fail to act as horizons.
However, if $\sigma=0$, $\cal{R}$ also becomes 0 for all values of $r$ and there is
no singularity at $r=m\pm M$. For $\sigma=0$, however, the solution reduces to the KN solution,
and the scalar field becomes trivial. So the only black hole solution in
this spacetime is KN black hole and hence this solution supports the
theorem for the nonexistence of a scalar hair.

\subsubsection{Second set}
Next we consider the case when $\phi$ is both function $R$ and $\theta$ in
the form $\phi=\alpha(R)+\beta(\theta)$. Here we have assumed the integration constants $\sigma$
and $\tau$ to be 0 in (A.5). Then from equation (A5) we get the solution for $\phi$
to be 
$$
\phi=\phi_{0}+\lambda\ln{[(e^{R}-{M^{2}\over{4}}e^{-R})\sin\theta]},
\eqno{(2.28)}$$
where $\phi_{0}$ and $\lambda$ are two constants. Here also we
consider $\phi_{0}=0$ without any loss of generality.
\par
Using the similar technique as before we find the line element for the EMS field in the
Boyer Lindquist coordinate as
$$
ds^{2}=dt^{2}-{2mr-e^{2}\over{r^{2}+a^{2}\cos^{2}\theta}}(dt+a\sin^{2}\theta d\phi)^{2}
-(r^{2}+a^{2}\cos^{2}\theta)\left\{(r^{2}-2mr+a^{2}+e^{2})\sin^2\theta\right\}^{\lambda^2\over{2}}
$$
$$
\left\{d\theta^{2}+{dr^{2}\over{r^{2}-2mr+a^{2}+e^{2}}}\right\}
-(r^{2}+a^{2})\sin^{2}\theta d\phi^{2},\eqno{(2.29)}$$
with the same solution for vector potentials $A_0$ and $A_3$. The scalar field,
when expressed in Boyer
Lindquist coordinates $(r,\theta)$, appears as
$$
\phi={\lambda\over{2}}\ln[(r^{2}-2mr+a^{2}+e^{2})\sin^2\theta].
\eqno{(2.30)}$$
This solution has metric singularities at $r=m\pm M$ surfaces, as
$g_{\theta\theta}$ goes to zero and $g_{11}$ goes to zero or infinity
corresponding to $\lambda>\sqrt{2}$ or $<\sqrt{2}$.
And like the previous case the scalar field $\phi$ diverges at these
surfaces. To ensure the nature of singularity we find the Ricci scalar
$$
{\cal{R}}=-\phi_{\alpha}\phi^{\alpha}
={\lambda^2\over{(r^2+a^2\cos^2\theta)}}[(r^2-2mr+a^2+e^2)\sin^2\theta]^{-\lambda^2\over{2}}$$
$$
\times\left\{{{(r-m)^2-M^2\cos^2\theta}\over{r^2-2mr+a^2+e^2}}\right\}.
\eqno{(2.31)}$$
For $\lambda=0$, $\cal{R}$ becomes 0 for all $r$ and for $\lambda\neq 0$
$\cal{R}$ diverges for $r=m\pm M$. So for $\lambda\neq 0$ these surfaces
$r=m\pm M$ become singular and hence fail to act as event horizons. As
for $\lambda=0$, $\cal{R}$ is finite and consequently the surfaces can act as 
event horizons. However, the scalar field becomes trivial in that case
and the metric reduces to the KN one. So this class of solutions also supports the
conjecture. 

\subsubsection{Third set}
We choose our third set of solution for $\phi$ from the general
solution (A6) where $\phi$ is separable in the product form of function
of $R$ and $\theta$. The solution, being the product of two infinite series,
we take the simplest choice $(n=1, ie, \lambda=2)$, given by(A16)  
$$
\phi=\sigma(e^R+{M^2\over{4}}e^{-R})\cos\theta,
\eqno{(2.32)}$$
where $\sigma$ is a  constant. 
\par
Adopting a similar technique we find the line element for EMS field in the
Boyer Lindquist form as 
$$
ds^{2}=dt^{2}-{2mr-e^{2}\over{r^{2}+a^{2}\cos^{2}\theta}}(dt+a\sin^{2}\theta d\phi)^{2}
-(r^{2}+a^{2}\cos^{2}\theta)\left\{e^{{-\sigma^2\over{2}}(r^{2}-2mr+a^{2}+e^{2})\sin^2\theta}\right\}
$$
$$
\left\{d\theta^{2}+{dr^{2}\over{r^{2}-2mr+a^{2}+e^{2}}}\right\}
-(r^{2}+a^{2})\sin^{2}\theta d\phi^{2},\eqno{(2.33)}$$ 
with the same solutions for the vector potentials. In the Boyer Lindquist
coordinate $(r,\theta)$ the scalar field looks like 
$$
\phi=\sigma(r-m)\cos\theta.
\eqno{(2.34)}$$
 It is quite transparent from the metric (2.33) that there is
singularity in $g_{11}$ at $r=m\pm M$ surfaces though the scalar field
remains finite at these surfaces. The expression for Ricci scalar is
$$
{\cal{R}}=-\phi_{\alpha}\phi^{\alpha}
=\sigma^2{(r-m)^2-M^2\cos^2\theta\over{(r^2+a^2\cos^2\theta)}}e^{{\sigma^2\over{2}}(r^{2}-2mr+a^{2}+e^{2})\sin^2\theta}.
\eqno{(2.35)}$$
 Unlike the other two cases for $\sigma\neq 0$ (equations (2.27)and
(2.31)), $\cal{R}$ remains
finite at $r=m\pm M$ surfaces. This implies that even for a non trivial scalar
field we have finite $\cal{R}$ at $r=m\pm M$ surfaces which, in turn,
indicates that these surfaces are no longer singular surfaces. In
fact the  Kretchman scalar $I$
$(I=R^{\mu\nu\alpha\beta}R_{\mu\nu\alpha\beta})$ also remains finite
everywhere including the surfaces $r=m\pm M$ surfaces for $\sigma\neq 0$.
The expression for $I$ is excluded from the text for the economy of space.
So these surfaces are not singular and act as event horizons to shield
the essential singularity. And as there is a nontrivial contribution of
scalar field this solution seems to
 contradict the `no scalar hair theorem'. But this solution
has two major defects. The solution is not 
asymptotically flat and the scalar field becomes infinite
for $r\rightarrow\infty$ limit, while the
energy due to scalar field has a finite contribution in that limit. So
although this solution has some nontrivial contribution for
$\phi$ on the horizon it could not really be considered as a serious
counter example due to its
pathological behaviour at $r\rightarrow \infty$.

\section{A conformal transformation and nonminimally coupled scalar fields}

\subsection{Conformal transformation}

The action for a very general scalar tensor theory
along with the Maxwell field is given by
$$
S[g_{\mu\nu},\phi,F_{\mu\nu}]=\int[f(\phi)R-h(\phi)g^{\mu\nu}\phi_{,\mu}\phi_{,\nu}
-F_{\mu\nu}F^{\mu\nu}]\sqrt{-g}d^{4}x,
\eqno{(3.1)}$$
with $f(\phi)>0$ and $h(\phi)>0$ where $g_{\mu\nu}, \phi$ and $F_{\mu\nu}$ are the metric tensor, the scalar field 
and the Maxwell field respectively. The scalar field is nonminimally coupled 
to gravity because of the term $f(\phi)$ in the action and the Newtonian constant 
G thus becomes a function of $\phi$ instead of being a constant. For different choices
of the functions $f(\phi)$ and $h(\phi)$, one obtains various scalar tensor
theories of gravitation. With a conformal transformation
$$
\bar{g}_{\mu\nu}=\Omega^{2}g_{\mu\nu},
\eqno{(3.2)}$$
where $\Omega^{2}=f(\phi),$
and by defining a new scalar field $\bar{\phi}$ in terms of $\phi$ as 
$$
\bar{\phi}(\phi)=\sqrt{2}\int^{\phi}_{\phi_{c}}d\xi\sqrt{{3\over{2}}({d\over{d\xi}}
\ln f(\xi))^{2}+{h(\xi)\over{f(\xi)}}},
\eqno{(3.3)}$$
the action (3.1) can be written in the form 
$$
\bar{S}[\bar{g}_{\mu\nu},\bar{\phi},\bar{F}_{\mu\nu}]=
\int[\bar{R}-{1\over{2}}\bar{g}_{\mu\nu}\bar{\phi_{,\mu}}\bar{\phi_{,\nu}}-
\bar{F^{2}}]\sqrt{-\bar{g}}d^{4}x.
\eqno{(3.4)}$$
Here $\phi_{c}$ is an arbitrary positive constant and the variables with an overhead
bar represent those in the transformed version. The action (3.4) clearly resembles that of a 
minimally coupled scalar field with a Maxwell field. So if the solution for this 
case is known, one can now easily find out the solutions for the corresponding 
nonminimally coupled scalar field cases by using the equations (3.2) and (3.3)
with proper choices for $f(\phi)$ and $h(\phi)$.
This type of conformal transformation has been used for a long time 
in the literature\cite{R12,R19,R24}.  
 Ref.\cite{R25}
represents a good set of references on this subject. 
\par
It deserves mention at this point\cite{R19*} that in case of dilaton gravity, obtained from the 
low energy limit of string theory, there is a coupling between Maxwell field and dilaton field in the action. This coupling makes dilaton gravity different from other scalar tensor theories described by action (3.1). 
Horne and Horowitz\cite{R32} found the black hole solution with dilaton hair 
for slow rotation in this theory.

\subsection{Some axisymmetric solutions in nonminimally coupled scalar tensor theories}
 
Amongst the three classes of solutions exhibited in section 2 each class of
solutions could be used  to generate the corresponding new solutions in
nonminimally coupled scalar tensor theories and then the
no scalar hair theorem will be verified against them. We cite the
examples in Brans-Dicke theory, although this method works for other
more general theories also.

\subsubsection{First set}
\noindent
The relevant action in BD theory\cite{R26} is 
$$
S=\int[\phi R-{\omega\over{\phi}}g^{\mu\nu}\phi_{,\mu}\phi_{,\nu}-F_{\mu\nu}F^{\mu\nu}]\sqrt{-g}d^{4}x.
\eqno{(3.5)}$$
Hence the conformal transformation for Brans Dicke field from the minimally coupled scalar field 
would be of the form 
$$
g_{\mu\nu}={1\over{\phi}}\bar{g}_{\mu\nu},
\eqno{(3.6)}$$
and
$$
\bar{\phi}(\phi)=\sqrt{(2\omega+3)}\ln{\phi\over{\phi_{c}}},
\eqno{(3.7)}$$
The solution for the Brans Dicke scalar field corresponding to (2.26) is
$$
\phi=\phi_{c}\left({r-(m+M)\over{r-(m-M)}}\right)^{\sigma\over{2M\sqrt{2\omega+3}}},
\eqno{(3.8)}$$
and the BDM metric components become 
$$
g_{\mu\nu}={1\over{\phi_{c}}}\left({r-(m-M)\over{r-(m+M)}}\right)^{\sigma\over{2M\sqrt{2\omega+3}}}\bar{g}_{\mu\nu}.
\eqno{(3.9)}$$
Thus the line element for the BDM metric in the Boyer Lindquist form is
$$
\phi_{c}ds^{2}=\left({r-(m-M)\over{r-(m+M)}}\right)^{\sigma\over{2M\sqrt{2\omega+3}}}
[dt^{2}-{2mr-e^{2}\over{r^{2}+a^{2}\cos^{2}\theta}}(dt+a\sin^{2}\theta d\phi)^{2}$$
$$
-(r^{2}+a^{2}\cos^{2}\theta)\left\{{r^{2}-2mr+a^{2}+e^{2}\over{(r-m)^{2}-M^{2}\cos^{2}\theta}}\right\}^{\sigma^{2}\over{2M^{2}}}
\left\{d\theta^{2}+{dr^{2}\over{r^{2}-2mr+a^{2}+e^{2}}}\right\}$$

$$\hspace{35mm}
-(r^{2}+a^{2})\sin^{2}\theta d\phi^{2}]\eqno{(3.10)}$$
with the solution for the vector potentials being the same as in EMS field. 
In the limit $r\rightarrow\infty$, the metric is  flat and the scalar 
field is constant. For this line element
the Ricci scalar takes the from
$$
{\cal{R}}={\omega\over{\phi^{2}}}g^{\mu\nu}\phi_{\mu}\phi_{\nu}
={\omega\sigma^{2}\over{2\omega+3}}\times{1\over{r^{2}+a^{2}\cos^{2}\theta}}\times
{[(r-m)^{2}-M^{2}\cos^{2}\theta]^{\sigma^{2}\over{2M^{2}}}\over{[r^{2}-2mr+a^{2}+e^{2}]^{1+{\sigma^{2}\over{2M^{2}}}}}}.
\eqno{(3.11)}$$
We find that the surfaces $r=m\pm M$ act as
physically singular surfaces for $\sigma\neq0$ as Ricci scalar $\cal{R}$ 
becomes infinitely large at these surfaces. 
And as there is no other horizon, these singularities are
naked. But if $\sigma=0$, the Ricci scalar remains finite and the scalar
field becomes trivial. So these surfaces are only coordinate singularities
and act as event horizon to shield the essential singularity. But 
the metric (3.10) reduces to the KN metric for $\sigma=0$. So like the minimally coupled
counterpart this set of solution for BDM spacetime also is in agreement with the no scalar hair
theorem as the only black hole solution is the KN solution which indeed
has no scalar hair. 

\subsubsection{Second set}
The solution for Brans Dicke scalar field in this case is
$$
\phi=\phi_c[(r^2-2mr+a^2+e^2)\sin^2\theta]^{\lambda\over{2\sqrt{2\omega+3}}},
\eqno{(3.12)}$$
and the line element in Brans Dicke Maxwell theory corresponding to the metric
(2.29) is
$$
\phi_c ds^2=[(r^2-2mr+a^2+e^2)\sin^2\theta]^{-\lambda\over{2\sqrt{2\omega+3}}}
[dt^{2}-{2mr-e^{2}\over{r^{2}+a^{2}\cos^{2}\theta}}(dt+a\sin^{2}\theta d\phi)^{2}$$
$$
-(r^{2}+a^{2}\cos^{2}\theta)\left\{(r^{2}-2mr+a^{2}+e^{2})\sin^2\theta\right\}^{\lambda^2\over{2}}
\left\{d\theta^{2}+{dr^{2}\over{r^{2}-2mr+a^{2}+e^{2}}}\right\}$$
$$
-(r^{2}+a^{2})\sin^{2}\theta d\phi^{2}].\eqno{(3.13)}$$
The expression for the Ricci scalar is
$$
{\cal{R}}={\omega\over{\phi^2}}g^{\mu\nu}\phi_{,\mu}\phi_{,\nu}$$
$$
={\lambda^2\omega\over{2\omega+3}}
{[(r^2-2mr+a^2+e^2)\sin^2\theta]^{-\lambda^2\over{2}}\over{(r^2+a^2\cos^2\theta)}}
\left\{{{(r-m)^2-M^2\cos^2\theta}\over{r^2-2mr+a^2+e^2}}\right\}.
\eqno{(3.14)}$$
It is quite clear that the surfaces $r=m\pm M$ are physically singular 
surfaces for non trivial scalar field i.e, for
$\lambda\neq 0$. For $\lambda=0$ these surfaces become event horizons with
finite Ricci scalar and trivial scalar field and the metric becomes KN. So
for this class of solution, like the previous case, we find that Brans Dicke
scalar hair is not compatible with the black hole. This solution can be
regarded as asymptotically zero curvature  solution since the
curvature (Ricci
scalar) becomes zero at $r\rightarrow\infty$.

\subsubsection {Third set}
Now we generate the third class of solutions corresponding to the
equations (2.33-2.35) which is fairly interesting in
the sense that it goes against the theorem.
For Brans Dicke theory the solution for the Brans-Dicke field in this case is
$$
\phi=\phi_c e^{{\sigma\over{\sqrt{2\omega+3}}}(r-m)\cos\theta}.
\eqno{(3.15)}$$
The corresponding line element of Brans Dicke Maxwell metric is
$$
\phi_c ds^2=e^{{-\sigma\over{\sqrt{2\omega+3}}}(r-m)\cos\theta}
[dt^{2}-{2mr-e^{2}\over{r^{2}+a^{2}\cos^{2}\theta}}(dt+a\sin^{2}\theta d\phi)^{2}$$
$$
-(r^{2}+a^{2}\cos^{2}\theta)\left\{e^{{-\sigma^2\over{2}}(r^{2}-2mr+a^{2}+e^{2})\sin^2\theta}\right\}
\left\{d\theta^{2}+{dr^{2}\over{r^{2}-2mr+a^{2}+e^{2}}}\right\}$$
$$
-(r^{2}+a^{2})\sin^{2}\theta d\phi^{2}].\eqno{(3.16)}$$ 
The expression for the Ricci scalar is
$$
{\cal{R}}={\omega\over{\phi^2}}g^{\mu\nu}\phi_{\mu}\phi^{\nu}$$
$$
={\sigma^2\omega\over{2\omega+3}}{(r-m)^2-M^2cos^2\theta\over{(r^2+a^2\cos^2\theta)}}
e^{{\sigma^2\over{2}}(r^{2}-2mr+a^{2}+e^{2})\sin^2\theta+{\sigma\over{\sqrt{2\omega+3}}}(r-m)\cos\theta}.
\eqno{(3.17)}$$
The expression for Ricci scalar reveals
that the surfaces $r=m\pm M$ are not physically singular but rather act as event
horizon with a nontrivial scalar field. But this solution,
like its minimally coupled counter part, has the defects of not being asymptotically
flat and having a divergent scalar field at infinite $r$. Thus, although
in this we have a horizon with nontrivial scalar field, it cannot be
taken seriously for its pathological asymptotic behaviour. 
\par
In all these three sets, we have examined some other non minimally
coupled scalar tensor theories\cite{R29,R30,R31}, where the BD parameter $\omega$ is a
function of the scalar field $\phi.$ In all these cases the results
are the same, i.e., for the first two sets there are no black holes with scalar hair and for the third 
the scalar hair can exist 
where the spacetime is not asymptotically flat. We do not include the
examples in the text for the economy of space.

\section{Discussion}
Although there are already a lot of results regarding the scalar hair
for spherical black holes, the axially symmetric black holes warrants
more investigations. The present work studies axisymmetric charged black holes
for various scalar tensor theories. All these solutions
are essentially the analogue of Kerr-Newman (KN) solutions in general
relativity and reduces to KN if the scalar field contribution is put
equal to zero. 
\par
For the first two sets, it is found that there is no regular
horizons if the scalar field exists. If, however, the scalar field is trivial 
$(\sigma=0)$, there are event horizons. But in the latter case,
$(\sigma=0)$ the metric is that of Kerr-Newman, and the geometry is endowed 
with only mass, electric charge and angular momentum.
\par
 For the third set of 
solutions, it appears that
the surfaces $r=m\pm M$ will act as event horizon as the scalar
curvatures are finite at those surfaces even with a non trivial scalar field. 
But these solutions are not asymptotically flat
and hence will produce curvature in the spacetime even at infinitely large
distances from the black hole.
\par
  
Thus the present investigations are in keeping with Bekenstein's
statement \cite{R35} that there is no
asymptotically flat, stationary and stable black hole solution in general
relativity and general scalar tensor theory which is endowed with a scalar
field. This also, in a way, is in agreement with  the Hawking theorem\cite{R36} 
which states that exterior of a stationary black hole is identical both 
in general relativity and Brans Dicke theory and this theorem can be 
extended to include a wide class of scalar tensor theories represented 
by action (3.1).
\par
The case of Kerr black hole analogue with the scalar fields can easily be
studied from the present work simply by setting the electric charge $e=0$. 
It is a trivial
matter to see that conclusions regarding the scalar hair will remain
exactly the same as in the case when the distribution has a nonzero charge.
\par
It deserves mention that the conformal transformation used in the
present work crucially depends on the
fact that $f(\phi)$ is positive. This may not be treated as a serious
restriction as in the weak field limit $f(\phi)$ gives the inverse of the
Newtonian constant $G$ and thus a negative $f(\phi)$ will indicate that
$G$ is negative. Anyway, for the sake of completeness the cases with
negative $f(\phi)$ should also be investigated. It should be noted  that
the solutions obtained by the generation techniques do actually solve the
relevant field equations.

\section{Acknowledgement}
One of us(S.S.) is thankful to the University Grants Commission, India for the
financial support.

\section{appendix}
We obtain the solution for the minimally coupled scalar field $\phi$ 
from equation (2.19) by assuming the separability condition. We first exhibit the solution when
$\phi$ is separable in summation form,
$$
\phi=\alpha(R)+\beta(\theta)\eqno{(A1)}$$
where $\alpha$ and $\beta$
are functions of $R$ and $\theta$ respectively.
Under such an assumption equation (2.19) attains the form
$$
\alpha_{RR}+{e^{R}+{M^{2}\over{4}}e^{-R}\over{e^{R}-{M^{2}\over{4}}e^{-R}}}
\alpha_{R}=-\{\beta_{\theta\theta}+{\cos\theta\over{\sin\theta}}\beta_{\theta}\}=\lambda,\eqno{(A2)}$$
where $\lambda$ is a separation constant.\\
 From (A2) it is quite easy to find the solution for $\alpha$ and $\beta$ as 
$$\alpha=\lambda\ln{(e^{R}-{M^{2}\over{4}}e^{-R})}+
{\sigma\over{M}}\ln\left({e^{R}-{M\over{2}}\over{e^{R}+{M\over{2}}}}\right)+const.,\eqno{(A3)}$$
and $$
\beta=\lambda\ln\sin\theta+\tau\ln\tan{\theta\over{2}}+const.,\eqno{(A4)}$$
where $\sigma$ and $\tau$ are integration constants. So 

$$\phi=\lambda\ln{[(e^{R}-{M^{2}\over{4}}e^{-R})\sin\theta]}+{\sigma\over{M}}\ln\left({e^{R}-{M\over{2}}\over{e^{R}+{M\over{2}}}}\right)+\tau\ln\tan{\theta\over{2}},\eqno{(A5)}$$
\\

Next we consider $\phi$ to be separable as a  product of functions of $R$ and $\theta$ as
$$\phi=\sigma A(R)B(\theta)\eqno{(A6)}$$
where $\sigma$ is constant.\\
With the form like (A8) equation (2.19) can be split into two equations,
$$A_{RR}+{e^{R}+{M^{2}\over{4}}e^{-R}\over{e^{R}-{M^{2}\over{4}}e^{-R}}}A_R-\lambda A=0\eqno{(A7)}$$
and $$B_{\theta\theta}+{\cos\theta\over{\sin\theta}}B_{\theta}+\lambda B=0,
\eqno{(A8)}$$
where $\lambda$ is the separation constant.\\
Now equation (A10) can be recast as 
$$
(1-X^2){d^2B\over{dX^2}}-2X{dB\over{dX}}+\lambda B=0\eqno{(A9)}$$
where $X=\cos\theta$. If $\lambda$ is taken as $n(n+1)$ where $n$ is
an integer, equation (A11) becomes the  Legendre differential equation of second order. Hence the solution of (A11) is given by the Legendre polynomial functions, i.e,
$$B(\cos\theta)=P_n (\cos\theta)
\eqno{(A10)}$$
where $$P_n (X)={1\over{2^n n!}}({d\over{dX}})^n(X^2-1)^n.$$
 Almost in a similar fashion equation (A9) is recast as
$$(M^2-Y^2){d^2A\over{dY^2}}-2Y{dA\over{dY}}+n(n+1) A=0,\eqno{(A11)}$$
where $Y={e^{R}+{M^{2}\over{4}}e^{-R}}$ and $\lambda=n(n+1)$.  
We find the series solution of this second order differential equation 
by substituting 
$$A=Y^k\sum_{l=0}^{\infty} a_l Y^l\hspace{3mm} (Frobenius\hspace{2mm} method)\eqno{(A12)}$$
The series solution for (A13) is 
$$A= a_0[1-{n(n+1)\over{2!}}({Y\over{M}})^2+{n(n+1)(n-2)(n+3)\over{4!}}({Y\over{M}})^4+....]
+ a_1[Y-{(n-1)(n+2)\over{3!}}{Y^3\over{M^2}}+........]\eqno{(A13)}$$
Now after normalisation $[S_n(M)=1]$ we find the generating function for the series 
$$S_n (Y)={1\over{(2M)^n n!}}({d\over{dY}})^n(Y^2-M^2)^n \eqno{(A14)}$$
This function generates similar polynomial as the Legendre ones 
except an extra factor of $M$ in the denominator.
$$S_0(Y)=1\hspace{5mm}P_0(X)=1$$
$$S_1(Y)={Y\over{M}}\hspace{5mm}P_1(X)=X$$
$$S_2(Y)={1\over{2M^2}}(3Y^2-M^2)\hspace{5mm}P_2(X)={1\over{2}}(3X^2-1)$$
$$S_3(Y)={1\over{2}}(5[{Y\over{M}}]^3-3{Y\over{M}})\hspace{5mm}P_3(X)={1\over{2}}(5X^3-3X)$$
So the solution for the scalar field is
$$\phi=\sigma A(R)B(\theta)=\sigma A(e^{R}+{M^{2}\over{4}}e^{-R})
B(\cos\theta)=\sigma A(Y)B(X)$$
$$=\sigma S_n(Y)P_n(X)=\sigma S_n(r-m)P_n(\cos\theta)\eqno{(A15)}$$
The simplest choice for $\phi$ from (A15) would be for $n=1$ 
(since for $n=0$, $\phi$ would be trivial), i.e,
$$\phi=\sigma S_1(Y)P_1(\cos\theta)=\sigma (e^{R}+{M^{2}\over{4}}e^{-R})
\cos\theta=\sigma(r-m)\cos\theta.\eqno{(A16)}$$

\end{document}